\documentclass{ws-p8-50x6-00}
\begin{document}

\title{ON NON HADRONIC ORIGIN OF HIGH ENERGY
 NEUTRINOS}

\author{H. ATHAR}

\address{Physics Division, National Center for
 Theoretical Sciences, Hsinchu 300, Taiwan and
 Institute of Physics, National Chiao Tung University, Hsinchu 300, Taiwan\\
 E-mail: athar@phys.cts.nthu.edu.tw}

\author{G.-L. LIN}

\address{
 Institute of Physics, National Chiao Tung University, Hsinchu 300, Taiwan\\
 E-mail: glin@cc.nctu.edu.tw}

\maketitle

\abstracts{Some of the  non hadronic interactions, such as the
$\eta$ resonance formation in the $\gamma \gamma $ interactions
and the muon pair production in the $e\gamma $ interactions, are
identified as possible source interactions for generating high
energy neutrinos in the cosmos.}

\section{Introduction}

At present\footnote{Talk given at First NCTS Workshop on
Astroparticle Physics, Kenting, Taiwan, 6-9 December, 2001.}, a
main motivation for high energy neutrino astronomy ($E_{\nu} \geq
10^{3}$ GeV) is that it may identify the role of hadronic
interactions taking place in cosmos \cite{kenting}. The hadronic
interactions mainly include the $p\gamma $ and $pp$ interactions.
These interactions produce unstable hadrons that decay into
neutrinos of all three flavors. There is a formation of $\Delta $
resonance in $p\gamma$ interactions, at center-of-mass energy,
$\sqrt{s}\sim m_{\Delta}$, that mainly decay into electron and
muon neutrinos~\cite{Greisen:1966jv}. In an astrophysical site for
these interactions, the protons are considered to be accelerated
up to a certain maximum energy and then interact with the photons
and other protons present in the vicinity of the source and those
present in the interstellar medium. Our galaxy and the earth
atmosphere are two examples of such astrophysical
sites~\cite{Athar:2001jw}.

Currently, the detectors taking data in the context of high energy
neutrinos are Antarctic Muon and Neutrino Detector Array (AMANDA)
at south pole~\cite{kenting} and the lake Baikal array in
Russia~\cite{Domogatsky:2001br}. These detectors are primarily
based on muon detection and are commonly referred to as high
energy neutrino telescopes. The other high energy neutrino
telescope under construction is the Astronomy with a Neutrino
Telescope and Abyss environmental RESearch (ANTARES)
project~\cite{Montaruli:2002gh}. These high energy neutrino
telescopes (envisage to) measure the showers and the charged
leptons produced mainly in the deep-inelastic neutrino-nucleon and
resonant (anti electron) neutrino-electron scatterings occurring
near or inside the high energy neutrino
telescopes~\cite{Athar:2002gs}. The later interaction can be used
to calibrate the incident neutrino energy in future high energy
neutrino telescopes. The Monopole, Astrophysics and Cosmic Ray
Observatory (MACRO) in Gran Sasso laboratory, Italy has also
recently reported its results for the high energy neutrino
searches~\cite{:2002ma}.  Given the present upper bounds on the
high energy neutrino flux from AMANDA (B 10) and Baikal detector,
the role of semi and non hadronic interactions becomes relevant.
We shall call the later interactions as purely electromagnetic
ones. Examples of these include $ep$  and $\gamma \gamma $
interactions respectively.

Upper bound from the AMANDA detector rule out some of the high
energy neutrino flux models based on hadronic interactions only.
However, several variants of these models can still possibly be
compatible with the high energy neutrino non observations. These
include, for instance, the direct pion production off the $\Delta
$ resonance in $p\gamma $ interactions.

 The absolute high energy
neutrino flux originating from the non hadronic interactions,
though expected to be small relative to that from hadronic
interactions, can be a good scale for future large high energy
neutrino telescopes such as IceCube. This will be a guaranteed
level of the high energy neutrino flux should the conventional
astrophysics explanation for observed high energy photon emission
from extra galactic astrophysical sources such as AGNs is correct
~\cite{science}. Here, only electromagnetic interactions are taken
into account for explaining the observations. Thus, the implicit
assumption of proton acceleration can be avoided. The discussion
that follows is also relevant in cases where the highest energy
cosmic rays, considered to be mainly protons, may not originate
from the GRBs which are the likely sources of high energy gamma
 rays~\cite{Scully:2000dr}.

This contribution is organized as follows. In Section 2, we
briefly discuss some essentials of purely electromagnetic
interactions possibly taking place in  astrophysical and
cosmological sites. In Section 3, we summarize the main points.

\section{Purely electromagnetic interactions}
\label{emt}

The non hadronic interactions  are defined to  have $e^{\pm}$ and
$\gamma $ in the initial state, rather than $p$ and $\gamma $.
Therefore, the possible interactions that may generate high energy
neutrinos include
%
%
\[
 \gamma \gamma \to \mu^{+}\mu^{-}, \,
 e\gamma \to \gamma^{'}\nu_{l}\bar{\nu}_{l}, \,
 e^{+}e^{-}\to \nu_{l}\bar{\nu}_{l}.
 \]
For comparison, note that for $\sqrt{s}\sim m_{\Delta}$, the cross
sections for these interactions are typically $\ll \mu$b. For
$\sqrt{s}\geq m_{\pi^{\pm}}$, other channels such as $e\gamma \to
e \pi^{+}\pi^{-}$ and $\gamma \gamma \to \pi^{+}\pi^{-}$ also
become available for high energy neutrino generation. The non
hadronic interactions also include the magnetic field induced
interactions such as $\gamma \gamma \to \nu_{l}\bar{\nu}_{l}$,
which will be briefly commented later in this Section.

A yet another possibility to generate high energy neutrinos in
purely electromagnetic interactions is through the formation and
decay of $\eta $ resonance into (charged) pions in $\gamma \gamma
$ interactions ($\gamma \gamma \to \eta \to
\pi^{+}\pi^{-}\pi^{0}$). Let us consider in some detail a simple
implication of this purely electromagnetic interaction in the
context of high energy photon
propagation~\cite{nocit,nocita,Protheroe:1995ft,Lee:1996fp,Poppe:1986dq}
and consequent high energy neutrino generation. The cross section
for this interaction is given by
%
%
\begin{equation}
\label{sigma}
 \sigma(\gamma \gamma \to \eta \to \pi^{+}\pi^{-}\pi^{0},s)
 =\frac{5.2 \Gamma^{2}_{\eta}}
 {(s-m_{\eta}^{2})^{2}+\Gamma^{2}_{\eta} m^{2}_{\eta}},
\end{equation}
where $\Gamma_{\eta}\sim 1.18 $ KeV and $m_{\eta}\sim 547 $ MeV,
so that $\Gamma_{\eta}/m_{\eta}\sim 10^{-6}$. The peak cross
section is $\sigma^{\rm res}(s=m^{2}_{\eta})\leq \, 3 $ mb. Let us
remark that $\sigma^{\rm res}(s=m^{2}_{\Delta})\leq \, 0.5 $ mb,
however here $\Gamma_{\Delta}/m_{\Delta}\sim 10^{-3}$. The
branching ratio~\cite{Groom:in} for the light unflavored $\eta $
meson to decay into  $\pi^{+}$, $\pi^{-}$ and $\pi^{0}$ is $\sim
23\%$.

A useful quantity in the context of high energy  photon
propagation is the average interaction length defined by
%
%
\begin{equation}
\label{length}
 l(E)^{-1}=\int^{\infty}_{m^{2}_{\eta}/4E} n_{\rm b}(\epsilon)
 \mbox{d}\epsilon
 \int^{-1}_{+1} \left(\frac{1-\mu }{2}\right)
 \sigma(s) \mbox{d}\mu ,
\end{equation}
with $s=2\epsilon E (1-\mu )$ and $\mu = \cos \theta$. The lower
limit of integration for $\epsilon $ corresponds to a head on
collision so that $\mu\, =\, -1$. Note that for a high energy
photon with energy  $E \sim $ GeV, the background photon energy is
$\epsilon\sim $ GeV to form the $\eta $ resonance. For
definiteness, let us consider the interaction between high energy
photons and the cosmic microwave background (CMB) radiation which
has the following number density
%
%
\begin{equation}
\label{cmb}
 n_{\rm b}(\epsilon)=
 \frac{1}{\pi^{2}}\frac{\epsilon^{2}}{\exp(\epsilon /T_{\rm b}(0))-1}.
\end{equation}
Here $T_{\rm b}(0)\simeq 2.74 $ K.  The other ubiquitous photon
backgrounds include the infrared and ultraviolet radiation,
considered to exist in the present universe, particularly after
galaxy formation epoch. For simplicity, we ignore their effects as
well as the effects of a possible magnetic field present in the
cosmos.
%
%
\begin{figure}[t]
\begin{center}
\psfig{figure=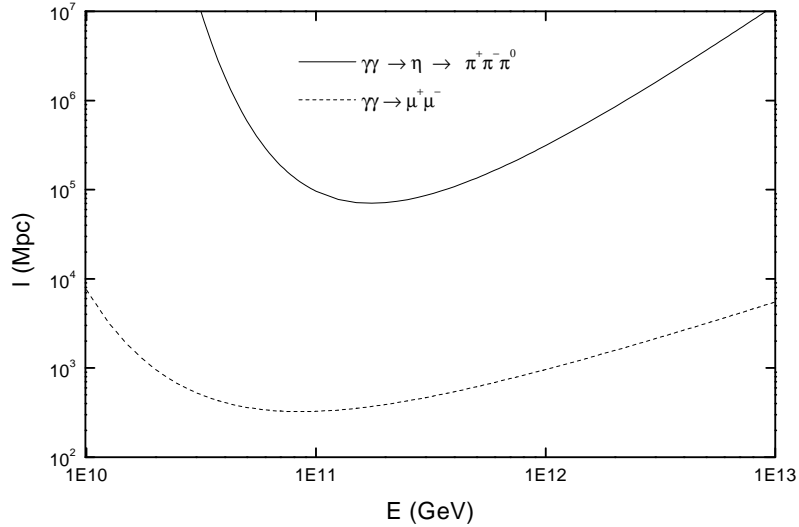,height=3.3in}
\end{center}
 \caption{The average interaction length using Eq. (\ref{length}) for high energy
 photons propagating in cosmic microwave background
 photon flux as a function of incident photon energy. \label{fig:one}}
\end{figure}
The high energy photons  are considered to originate from an
astrophysical or cosmological site within the present horizon
length $cH^{-1}(0)\sim 3\cdot 10^{3}$ Mpc (where 1 pc $\sim 3\cdot
10^{18}$ cm), as we take $H(0)\sim 65$ km s$^{-1}$ Mpc$^{-1}$.

Substituting Eq. (\ref{sigma}) and Eq. (\ref{cmb}) into Eq.
(\ref{length}), the two integrations can be carried out easily
 under the narrow width assumption,  such that for $\sqrt{s}\simeq
 m_{\eta}$, we obtain
%
%
\begin{equation}
\label{narrow}
 l(E)\simeq  \frac{8\pi E^{2}}{5.2 \Gamma_{\eta} m_{\eta} T}
 \left\{\ln\left(\frac{\exp(m^{2}_{\eta}/4ET)}
 {\exp(m^{2}_{\eta}/4ET-1)}\right)\right\}^{-1}.
\end{equation}
For $4ET\simeq m^{2}_{\eta}$, i.e., $E\sim 3\cdot 10^{11}$ GeV, we
note that $l\sim 10^4 \, {\rm Mpc} > cH^{-1}(0)$.  A  such single
interaction give a total of 6 neutrinos.  For comparison, we
display in Fig. \ref{fig:one}, the $l_{\eta}\equiv l(\gamma \gamma
\to \eta \to \pi^{+}\pi^{-}\pi^{0})$  along with the more familiar
relevant $l$, namely for $\gamma \gamma \to \mu^{+}\mu^{-}$. From
the figure, we note that the Eq. (\ref{narrow}) is a quite good
approximation to obtain $l$ in resonance and that $l_{\eta}\gg
l_{\mu^{+}\mu^{-}}$ for same $E$. In general, this observation may
also have some relevance for high energy photon propagation in a
dense photon background with relatively narrow background photon
flux spectrum such as those arising in some astrophysical sites in
the context of high energy neutrino generation.

In the limit $4ET\ll m^{2}_{\eta }$, we obtain
%
%
\begin{equation}
\label{limit1}
  l(E)\simeq \frac{8 \pi E^{2}}{5.2 \Gamma_{\eta} m_{\eta} T}
  \exp \left(\frac{m^{2}_{\eta}}{4ET}\right),
\end{equation}
whereas, in the opposite limit, namely when $4ET\gg m^{2}_{\eta
}$, we obtain
%
%
\begin{equation}
\label{limit2}
  l(E)\simeq \frac{8 \pi E^{2}}{5.2 \Gamma m_{\eta} T}
  \left\{\ln\left(\frac{4ET}{m^{2}_{\eta}}\right)\right\}^{-1}.
\end{equation}
In the two limiting cases,  $l(E)\gg cH_{0}^{-1}$. Let us further
remark that although  $\sigma^{\rm res}(s=m^{2}_{\eta})\gg
\sigma^{\rm res}(s=m^{2}_{\Delta})$, however $l^{\rm
res}_{\eta}\gg l^{\rm res}_{\Delta}$ because of rather narrow
$\eta$ width.

In the presence of an external magnetic field, we note that the
cross section for $\gamma\gamma\to \nu\bar{\nu}$ is significantly
enhanced~\cite{Shaisultanov:1997bc} with respect to its value in
the vacuum. However, such an enhancement is still insufficient for
this process to be presently relevant for high energy neutrino
generation. For comparison with $\gamma\gamma\to \mu^{+}\mu^{-}$,
it is found that~\cite{Chyi:1999wy},  for $B=10^{12}$ G,
$\sigma(\gamma\gamma\to \nu\bar{\nu})\approx 10^{-49} {\rm cm}^2$
for $s \, {\buildrel
>\over {_{\sim}}}\,  4 \, m_{\mu}^{2}$. This cross section scales
as $B^2$ for $B< B_c\approx 4\cdot 10^{13}$ G.

\subsection{Astrophysical sites}

Presently, there exists no model to estimate the high energy
neutrino flux in purely electromagnetic interactions taking place
in sources of highest energy gamma rays  such as the AGNs and the
GRBs. To make an order of magnitude estimate, we assume that the
above astrophysical sites can accelerate electrons to energies
greater than the observed gamma ray energies. As these electrons
undergo inverse Compton scattering, the up-scattered high energy
photons are produced. The scatterings of high energy photons over
the ambient photon fields present in the vicinity of the AGNs or
GRBs may lead to the $\mu^{+}\mu^{-}$ final state or three-pion
final state through the $\eta $ resonance.

Phenomenologically speaking, the resulting (relative) high energy
neutrino flux can be parameterized as $\phi^{\gamma
\gamma}_{\nu}(E_{\nu})\sim P \phi^{p \gamma}_{\nu}(E_{\nu})$,
where the probability function $P$ depends on the ratio of high
energy photon/electron flux associated with a specific
astrophysical site to the corresponding high energy proton flux on
the same site. The function $P$ certainly also depends on the
ratio of neutrino production cross sections between two
mechanisms. Finally it also depends on the magnetic field strength
on the site, which are relevant for the acceleration of charged
particles. A {\em diffuse} non hadronic high energy neutrino flux
with a representative $P$ $\sim 10^{-4}-10^{-5}$ can in principle
be measurable by future large high energy neutrino telescopes such
as IceCube.

\subsection{Cosmological sites}

Topological defects formed in the early epochs may play some role
in the latter epochs of the expanding universe. The cosmological
and astrophysical aspects of topological defects are the density
or metric perturbations that they may generate, particularly in
the epoch of large scale structure formation in the expanding
universe~\cite{Durrer:2001cg}.

The associated particle physics aspect is the possible release of
large amount of energy trapped inside these topological defects in
the form of gauge bosons. These gauge bosons subsequently decay
into known hadrons and leptons. Assuming that (some fraction of)
the topological defects are formed in the early epochs of the
expanding universe and thus contain a large amount of energy, it
becomes possible to explain at least some features of the observed
highest energy cosmic rays. For this scenario to work, the
observed highest energy cosmic rays have to be dominantly the
photons. In this (conventional) scenario, the high energy neutrino
flux is generated from the decay of charged
pions~\cite{Sigl:2001th}.

Here, we discuss a class of topological defects in which high
energy neutrino flux generation was postulated to originate in the
electromagnetic cascade rather than in charged-pion decays which
result from the hadronizations of initial jets produced in the
decays of GUT-scale heavy bosons~\cite{Kusenko:2000sq}. This class
of sources for ultrahigh energy photons is assumed to be active
before the galaxy formation epoch. This
corresponds~\cite{Peebles:xt} to a red shift, $z\, \geq 5$. Thus,
for $z\, \geq 5$, the effects of galactic magnetic field as well
as the infrared and ultraviolet photon backgrounds can be
neglected. Consequently, CMB photon flux is the only important
photon background. A search for high energy neutrinos can provide
some useful information about the existence of this class of
topological defects in the expanding universe. At  high red shift,
the $\gamma \gamma $ interactions between the energetic and
background photons can produce muons (and charged pions) whose
decay generate high energy neutrinos. Note that, at high red
shift, $T_{\rm b}(z)= (1+z) T_{\rm b}(0)$, whereas $n_{\rm
b}(z)=(1+z)^{3}n_{\rm b}(0)$. For $E> E_{\rm th}(z)$, where
$E_{\rm th}=10^{11} {\rm GeV}/(1+z)$, the $\gamma \gamma_{\rm
b}\to \mu^{+}\mu^{-}$ is most relevant for high energy neutrino
generation. The $\lambda \equiv \lambda (\gamma \gamma_{\rm b} \to
\mu^{+}\mu^{-}$) obtainable using Eq. (\ref{length}) is less than
the horizon length, $cH(z)^{-1}$ for $5 < z < 10$. With an
invariant mass just above the threshold, the purely
electromagnetic interaction $\gamma \gamma_{\rm b}\to
\mu^{+}\mu^{-}$ also has a shorter interaction length than the
energy attenuation length in the electromagnetic cascade dictated
by $\gamma \gamma_{\rm b}\to e^{+}e^{-}$.

Under the assumption that muons decay before interacting, the high
energy neutrino flux can be calculated as
%
%
\begin{equation}
\label{cosmoflux}
 \phi^{\gamma \gamma }_{\nu}(E_{\nu})= \int^{z_{max}}_{z_{min}} \int^{E_{max}}_{E_{\nu}}
 {\mbox d}z\,  {\mbox d}E\,  \phi_{\gamma}\, (z, E)\,  f(z, E)
 \frac{\mbox{d}n_{\gamma \gamma \to \nu+X}}   {\mbox{d}E_{\nu}}.
\end{equation}
Here $\phi_{\gamma}(z, E)$ parameterizes the high energy photon
flux from the topological defect. Typically, it is normalized by
assuming that the high energy photons produced by topological
defects at the high red shift are dominantly responsible for the
observed high energy photon flux and/or the observed highest
energy cosmic rays. The function $f(z, E)\equiv cH(z)^{-1}/\lambda
$ gives the number of $\gamma \gamma $ interactions within the
horizon length. The ${\rm d}n/{\rm d}E\equiv \sigma^{-1}{\rm
d}\sigma/{\rm d}E$ is the neutrino-energy distribution in $\gamma
\gamma $ interactions. The integration limits follow from the
above discussion. The $\phi^{\gamma \gamma }_{\nu}$ peaks at
$E_{\nu}\simeq E_{\mu}/3\simeq E_{\rm th}/3(1+z)\simeq 10^{11} \,
{\rm GeV}/3(1+z)^{2}\simeq 10^{8}$ GeV. The $\eta$ resonance
formation can also contribute to $\phi^{\gamma \gamma }_{\nu}$. It
is a possibility to produce high energy neutrinos through non
hadronic interactions in a cosmological setting.

The electromagnetic cascade that generate high energy neutrinos
from muon decays in $\gamma \gamma $ interactions contains roughly
equal number of photons and electrons. In
Ref~\cite{Kusenko:2000fk}, it was suggested that, for this class
of topological defects that produce ultrahigh energy photons at
the high $z$, the muon pair production (MPP) in $e^{-}\gamma \to
e^{-}\mu^{+}\mu^{-}$ dominates over the triplet pair production
(TPP) in $e^{-}\gamma \to e^{-}e^{+}e^{-}$ for 5 m$^{2}_{\mu} \leq
s\leq$ 20 m$^{2}_{\mu}$ in the electromagnetic cascade, thus
enabling the MPP process to be an efficient mechanism for
generating high energy neutrinos at the high $z$. The electrons in
the final state of the above processes are considered as
originating from the electromagnetic cascade generated by the
ultrahigh energy photons scattering over the  CMB photons present
at the high red shift. This conclusion was based upon the value of
the ratio $R$ defined as $R\simeq
\sigma_{MPP}/\eta_{TPP}\sigma_{TPP}$, where $\eta_{TPP}$ is the
inelasticity for the TPP process. The $\eta_{TPP}$ is basically
the average fraction of the incident energy carried by the final
state positron. The original estimate of Ref.
\cite{Kusenko:2000fk} gives $R\simeq 10^{2}$, which favors the MPP
process as the dominating high energy  neutrino generating
process. Namely, the electron energy attenuation length due to the
TPP process is much longer than the interaction length of the MPP
process because $\sigma_{MPP}\simeq (0.1-1) $ mb. However, by an
explicit calculation~\cite{Athar:2001sm}, instead it was found
that $\sigma_{MPP}<1\, \mu $b for $s\geq 5$ m$^{2}_{\mu}$, thus
yielding $R<1$. In particular,
%
%
\begin{equation}
\label{MPP}
 \sigma_{MPP}(s)= \left \{
           \begin{array}{ll}
          4\cdot 10^{-3} \, \mu \mbox{b} & {\rm for}\, \,  s= 4\,  {\mbox m}^{2}_{\mu}, \\
          1\cdot 10^{-1} \, \mu \mbox{b} & {\rm for}\, \,  s= 20\,  {\mbox
          m}^{2}_{\mu}.
        \end{array}
      \right .
\end{equation}
Therefore, MPP can not be a dominating process for generating high
energy neutrinos. We note that the equivalent photon approximation
was used in this work to calculate the leading-order contribution
to $\sigma_{MPP}(s)$.

In summary, in an electromagnetic cascade generated by ultrahigh
energy photons scattering over the CMB photons at the high red
shift, the $\gamma \gamma \to \mu^{+}\mu^{-}$ can in principle
produce high energy neutrinos, typically for $5 < z < 10$, through
the muon decays. On the other hand, the process $e\gamma \to
e^{-}\mu^{+}\mu^{-}$ occurring as the next round of interactions
in the same electromagnetic cascade can not produce the high
energy neutrinos.

\section{Conclusions}

Possibilities of high energy neutrino generations in two of the
non hadronic interactions, namely $\gamma \gamma $ and $e\gamma $
reactions are briefly discussed. In the first interaction, the
formation and decay of the $\eta$ resonance in addition to the
muon pair production may have some implications for high energy
neutrino generation. Model dependent analysis is needed to further
quantify the high energy neutrino generation in non hadronic
interactions.

\section*{Acknowledgments}
 HA thanks Physics Division of
National Center for Theoretical Sciences for financial support.
GLL is supported by the National Science Council of Taiwan under
the grant number NSC90-2112-M009-023.

\end{document}